\documentclass[aps, prb, showpacs, twocolumn, amsmath, letterpaper]{revtex4-2}

\usepackage{graphics}
\usepackage{graphicx}

\usepackage{amssymb}
\usepackage{bm}

\usepackage[english]{babel}

\usepackage{tabularx}
\usepackage{booktabs}

\usepackage{amsmath}
\usepackage{graphicx}
\usepackage{wrapfig}

\usepackage{adjustbox}
\usepackage{mathtools}  

\usepackage[colorlinks=true, allcolors=blue]{hyperref}

\begin{document}

\title{Restoring a Missing Meta-Symmetry of Quantum Mechanics}

\author{Sheng Ran}

\affiliation{Department of Physics, Washington University in St. Louis, St. Louis, MO 63130, USA}

\date{\today}

\begin{abstract}

In conventional quantum mechanics, all unitary evolution takes place within the space–time Hilbert space $\mathcal H_{xt}=L^2(\mathcal M_{xt})$, with time as the sole evolution parameter. The momentum–energy representation $\phi(k,E)$ is treated merely as a Fourier re-expression of the same state—kinematically equivalent but dynamically inert.
Here we restore the fundamental symmetry between the conjugate pairs $(x,t)$ and $(k,E)$ by extending the quantum theory to an enlarged Hilbert space $\mathcal H_{\text{total}} = \mathcal H_{xt} \oplus \mathcal H_{kE}$, within which the momentum–energy sector $\mathcal H_{kE}=L^2(\mathcal M_{kE})$ carries its own autonomous unitary evolution generated by a self-adjoint operator $\hat{\mathcal T}$. The resulting structure establishes a meta-symmetry: a symmetry between two conjugate dynamical projections of a single global quantum state. It produces a dual-manifold geometry in which each domain is locally complete yet globally open, with divergent limits in one mapping onto extended regions in the other. Remarkably, the dual-manifold symmetry alone reproduces both the uniform dark-energy background and the exponential boundary mapping near black-hole horizons that underlies Hawking radiation. This framework thus opens a quantum-theoretic route to cosmological phenomena that are ordinarily treated within general relativity.

\end{abstract}

\maketitle{}

%Symmetry underlies all physical laws. From rotational invariance in classical mechanics to Lorentz symmetry in relativity and gauge symmetry in quantum field theory, each symmetry principle defines a class of conservation laws and governs how nature organizes its dynamics. The discovery of a new symmetry—or the breaking of an existing one—has repeatedly marked the turning points of modern physics.

The foundation of quantum mechanics rests on a single profound statement: matter is fundamentally wave-like. The dynamics of this wave are governed by the Schrödinger equation \cite{Schroedinger1926}:
\[
i\hbar\frac{\partial \psi}{\partial t} = \hat{H}\psi.
\]
This equation places the wavefunction $\psi(x,t)$ on the space–time manifold $\mathcal{M}_{xt}$, and prescribes its evolution along the time coordinate $t$.

However, the structure of the plane wave,
\[
A e^{i(kx - Et/\hbar)},
\]
suggests an underlying symmetry between the conjugate coordinates $(x,t)$ and $(k,E)$ that the Schrödinger equation does not fully realize.
Mathematically, the two domains are related by a unitary Fourier transform:
\[
\phi(k,E)=\int dxdt\psi(x,t)e^{-i(kx-Et/\hbar)}.
\]
which defines an isomorphism between the associated function spaces,
\[
\mathcal F: L^2(\mathcal M_{xt})\rightarrow L^2(\mathcal M_{kE}).
\]
This is a representation symmetry: both descriptions encode the same physical state.

Yet physically this symmetry remains incomplete. In standard quantum mechanics, all observable dynamics unfold exclusively within $\mathcal{M}_{xt}$: even when written as $\phi(k,E)$, the wavefunction still evolves only in $t$. The momentum–energy representation does not possess its own generator of evolution, and no physical process unfolds in the variable $E$. From a symmetry viewpoint, nothing forbids such dynamics—there is simply no equation governing them.

The presence of momentum–energy waves that evolve in $E$ is not an artificial addition but an unavoidable consequence of quantum localization. A wavepacket that is localized in space or time—i.e., sharply peaked in $x$ or in $t$—necessarily corresponds to a delocalized, oscillatory structure in its conjugate variables $k$ or $E$. Under ordinary conditions this sector plays no dynamical role, but in extreme regimes—where spatial and temporal localization approach their fundamental limits and the smooth structure of $\mathcal M_{xt}$ breaks down—the momentum–energy manifold $\mathcal M_{kE}$ becomes the natural substrate for a coherent quantum wave. In such cases, the state may manifest as a genuine, extended excitation on $\mathcal M_{kE}$, governed by evolution in $E$ rather than $t$.

In this sense, the conventional assumption that the wavefunction lives solely on $\mathcal M_{xt}$ and evolves only in time captures only one side of a deeper conjugate structure. A century after Schrödinger’s equation, we propose to restore the missing dynamical symmetry between the dual manifolds $\mathcal M_{xt}$ and $\mathcal M_{kE}$. This is a meta-symmetry, because it is not a symmetry within a single dynamical system, but a symmetry between two autonomous evolution laws—one governed by time and the Hamiltonian, the other by energy and its conjugate generator. Mathematically it interchanges $(t,\hat{H})$ with $(E,\hat{\mathcal T})$, and physically it appears as a duality between the two dynamical sectors of a single quantum state.
This restored symmetry provides a more complete and balanced description of quantum waves and opens a new route to phenomena that traditionally require general relativity for their explanation.

\section{Dynamical framework on the momentum–energy manifold}

We first formulate the dynamics of a wave defined primarily on the momentum–energy manifold, $\mathcal{M}_{kE}$, whose intrinsic coordinates are $(k,E)$. Within this setting, the space–time parameters $(x,t)$ emerge not as fundamental coordinates but as Fourier-conjugate variables. Fourier duality between the coordinate pairs $(x,k)$ and $(t,E)$ defines the canonical operators on $\mathcal{M}_{kE}$: 
\[
\hat X=-i\partial_k,\qquad \hat{\mathcal T}=i\hbar\partial_E,
\]
so that on plane waves $\phi_{x,t}(k,E)=Ae^{i(kx-Et/\hbar)}$,
\[
\hat X \phi_{x,t}=x\phi_{x,t},\qquad \hat{\mathcal T}\phi_{x,t}=t\phi_{x,t}.
\]
Thus $x$ and $t$ appear as emergent conjugate parameters within $\mathcal{M}_{kE}$, generated by gradients in $k$ and $E$. This relation is purely kinematic, following directly from Fourier symmetry.

To elevate this duality to a dynamical symmetry, we postulate that states on $\mathcal{M}_{kE}$ evolve unitarily with respect to the energy coordinate $E$, in direct analogy with time evolution in ordinary quantum mechanics:
\[
\begin{split}
  \phi(k,E+\Delta E)=U(\Delta E)\phi(k,E), 
\end{split}
\]
where ${U(\Delta E)}_{\Delta E\in\mathbb R}$ forms a strongly continuous one-parameter unitary group,
\[
U(\Delta E_1+\Delta E_2)=U(\Delta E_1)U(\Delta E_2).
\]
By Stone’s theorem~\cite{Stone1930,Stone1932}, there exists a self-adjoint generator $\tilde H$ such that
\[
\begin{split}
 U(\Delta E)=e^{-i\tilde H\Delta E/\hbar}.
\end{split}
\]
Expanding to first order in $\Delta E$ then yields the Schrödinger-type evolution equation
\[
i\hbar\partial_E \phi(k,E)=\tilde H\phi(k,E).
\]
This conjugate Schrödinger equation is not a Fourier re-expression of the standard time-evolution equation. Rather, it represents an independent, complementary dynamical sector, completing the two-sided symmetry between the space–time and momentum–energy manifolds. Since the operator that generates $E$-translations is precisely $i\hbar\partial_E$, we identify
\[
\tilde H = \hat{\mathcal T}.
\]

We now constrain $\hat{\mathcal T}$ by the same physical principles used in deriving ordinary Hamiltonians:

1. $k$-translation invariance (homogeneity of $\mathcal{M}_{kE}$).
   Let $K(a)=e^{-i a \hat X}$ be translations in $k$: $K(a)\phi(k)=\phi(k+a)$.
   Homogeneity requires $[\hat{\mathcal T},K(a)]=0$ for all $a$. By the bicommutant argument, $\hat{\mathcal T}$ must be a function of $\hat X$ only:
   \[
   \hat{\mathcal T}=f(\hat X).
   \]

2. Parity (isotropy) in $k$.
   Under $k\mapsto -k$ we have $\hat X\mapsto -\hat X$. Isotropy implies $\hat{\mathcal T}$ is even:
   \[
   \hat{\mathcal T}=f(\hat X^2).
   \]

3. Locality.
   Requiring $\hat{\mathcal T}$ to act locally on $\phi(k,E)$ selects differential operators of finite order in $\hat X=-i\partial_k$. The lowest-order even form is quadratic:
   \[
   \hat{\mathcal T}=\alpha\hat X^2+V_{0},\ 
   \]
   with real constants $\alpha>0$ setting the stiffness of $E$-evolution and $V_0$ a reference level.

Putting together yields the minimal dynamics on $\mathcal{M}_{kE}$:
\[
\ i\hbar\partial_E \phi(k,E)=\Big(-\alpha\partial_k^2+V_0\Big)\phi(k,E).
\]
Higher-order even terms (e.g. $\beta\partial_k^4)$ encode dispersive corrections. Gauge-like couplings and curvature of $\mathcal{M}_{kE}$ can be incorporated via minimal substitution $\hat X\to\hat X-A_k(k)$ and covariantization $\partial_k\to\nabla_k$.

Having endowed $\mathcal{M}_{kE}$ with its own dynamics, we must clarify what this implies for the Hilbert-space structure. In standard quantum theory a single physical state $|\Psi\rangle$ is represented in a single Hilbert space $\mathcal H=L^2(\mathcal M_{xt})$. The $(x, t)$ and $(k, E)$ descriptions are simply two representations of the same state, related by the unitary Fourier transform. Although the function spaces $L^2(\mathcal M_{xt})$ and $L^2(\mathcal M_{kE})$ are isomorphic, only the $(x, t)$ representation carries genuine dynamics: every representation of the state evolves according to the same time-evolution law $i\hbar\partial_t\psi = H\psi$.

The structure changes once we promote $\mathcal M_{kE}$ to a fully dynamical manifold. Allowing the wavefunction to evolve in the energy variable introduces a second autonomous one-parameter unitary evolution, not derivable from the usual time dynamics. In this step the function space $L^2(\mathcal M_{kE})$ becomes a genuine dynamical Hilbert space. Concretely, we have
\[
\mathcal H_{xt}=L^2(\mathcal M_{xt}),\qquad
\mathcal H_{kE}=L^2(\mathcal M_{kE}),
\]
with the standard (kinematical)Fourier isomorphism
\[
\mathcal F:\mathcal H_{xt}\xrightarrow{\sim}\mathcal H_{kE},
\]
which, however, no longer represents a dynamical equivalence once $\mathcal H_{kE}$ carries its own autonomous evolution.

\begin{table*}
\centering
\caption{Structural comparison between standard quantum mechanics and the dual-manifold framework.}
\begin{tabular}{lll}
\toprule
\textbf{Feature} & \textbf{Standard QM} & \textbf{Dual-Manifold Dynamics} \\
\midrule

\textbf{Wavefunction} 
& $|\Psi\rangle$ with representations $\psi(x,t)$, $\phi(k,E)$
& Same  \\[3pt]

\textbf{Physical Manifold} 
& $\mathcal M_{xt}$
& $\mathcal M_{xt}$ and $\mathcal M_{kE}$ \\[3pt]

\textbf{Evolution in $t$} 
& Yes 
& Yes \\[3pt]

\textbf{Evolution in $E$} 
& No 
& Yes \\[3pt]

\textbf{Evolution generators}
& $\hat H$ only
& $\hat H$ and $\hat{\mathcal T}$ \\[3pt]

\textbf{Hilbert-space structure}
& $\mathcal H = L^2(\mathcal M_{xt})$
& Enlarged $\mathcal H_{\text{total}} = \mathcal H_{xt} \oplus \mathcal H_{kE}$ \\[3pt]

\textbf{Representation symmetry}
& $(x,t) \leftrightarrow (k,E)$
& $(x,t) \leftrightarrow (k,E)$ \\[3pt]

\textbf{Meta symmetry }
& No 
& $t \leftrightarrow E,\; \hat H \leftrightarrow \hat{\mathcal T}$\\

\bottomrule
\end{tabular}
\end{table*}

Because each space now carries its own strongly continuous one-parameter unitary group—$U(\Delta t)=e^{-iH\Delta t/\hbar}$ on $\mathcal H_{xt}$ and $U(\Delta E)=e^{-i\tilde H\Delta E/\hbar}$ on $\mathcal H_{kE}$—the natural structure is that of an enlarged Hilbert space containing two dynamical subspaces:
\[
\mathcal H_{\text{total}} = \mathcal H_{xt} \oplus \mathcal H_{kE}.
\]
The physical state remains single; it may be represented on either manifold, but the two representations now belong to distinct dynamical sectors rather than two views of the same evolution.

Crucially, the generators $H$ and $\hat{\mathcal T}$ are not related by Fourier conjugation,
\[
\hat{\mathcal T} \neq \mathcal F H\mathcal F^{-1}.
\]
Here the $E$-evolution is introduced as an independent dynamical ingredient, and the Fourier map connects only the representations of the state—not the generators governing their evolution. This independence is precisely what makes $\mathcal H_{xt}$ and $\mathcal H_{kE}$ two dynamical Hilbert subspaces of a single enlarged space, rather than two coordinate views of the same dynamical system.

In this sense the dual-manifold framework restores a symmetry that is absent in standard quantum theory:
\[
t \leftrightarrow E,\qquad
\hat{H} \leftrightarrow \hat{\mathcal T}.
\]
At the level of dynamical systems, this interchange acts not as a conventional representation symmetry but as a duality, relating two autonomous self-adjoint evolutions defined on the conjugate manifolds $\mathcal M_{xt}$ and $\mathcal M_{kE}$. Thus the theory realizes a meta-symmetry—mathematically a symmetry of the enlarged Hilbert-space structure, physically manifested as a duality between two independent dynamical sectors of the same quantum state.

This also explains, without contradiction, why a time operator is obstructed in the usual picture but appears naturally here. In the standard $t$-evolution framework on $\mathcal H_{xt}$, the Pauli argument forbids a self-adjoint $\hat T$ with $[\hat T,\hat{H}]=i\hbar$ when $\hat{H}$ is bounded below ~\cite{Pauli1980}. In our energy-evolution framework on $\mathcal H_{kE}$, the evolution parameter is $E$; the corresponding generator $\hat{\mathcal T}=i\hbar\partial_E$ is self-adjoint on its natural domain. Hence the appearance of a time operator is not a violation of Pauli’s objection—it is a consequence of working in the enlarged Hilbert space.

\section{Projections Between Dual Manifolds}

Having restored the dual structure between the space–time manifold and the momentum–energy manifold, we now examine the projections between the two manifolds.
Our inability to perceive the conjugate domains arises from the fact that waves native to $\mathcal M_{kE}$ do not propagate within $\mathcal M_{xt}$: a fully extended state in momentum–energy space projects to a pointlike, non-propagating configuration in space–time, and conversely. Thus information does not flow dynamically between the two domains; it can only appear in one manifold as the projection of a state that lives primarily in the other. 

In what follows, we show that two major cosmological phenomena emerge naturally from this projection structure: (1) a stationary background on $\mathcal M_{kE}$ projects into $\mathcal M_{xt}$ as a uniform and isotropic energy density—dark energy;
and (2) a geometric correspondence between extreme limits of the two manifolds explains the behavior of waves near black-hole horizons.

\subsection{Dark Energy from Pure Fourier Projection}

We begin with the simplest situation: the two manifolds are connected only by the purely unitary Fourier correspondence, with no physical coupling introduced. Consider a stationary background of quantum modes living in the momentum–energy manifold $\mathcal M_{kE}$, written in its most general decomposition into amplitude and phase,
\[
\phi(\mathbf k,E)=A(\mathbf k,E)e^{i\alpha(\mathbf k,E)}.
\]
%Such a representation involves no assumption; it is merely the polar decomposition of a complex wave amplitude.

Because observers in $\mathcal M_{xt}$ have no access to the microscopic phase information stored in $\mathcal M_{kE}$, the background must be treated as a random-phase ensemble, a standard assumption in wave kinetics, quantum statistical mechanics, and cosmology:
\[
\Big\langle e^{i[\alpha(\mathbf k,E)-\alpha(\mathbf k',E')]}\Big\rangle
\simeq \delta^{(3)}(\mathbf k-\mathbf k')\delta(E-E'),
\]
Projecting this ensemble through the purely unitary Fourier map,
\[
\psi(\mathbf x,t)
= \int d^3k dE
e^{i(\mathbf k\cdot\mathbf x-Et/\hbar)}\phi(\mathbf k,E),
\]
and averaging over the inaccessible phases yields the diagonal expectation
\[
\langle|\psi(\mathbf x,t)|^2\rangle
= \int d^3k dE |A(\mathbf k,E)|^2
\equiv \rho_\Lambda,
\]
which is spatially and temporally constant.

Thus, without invoking any dynamical coupling, the background mode distribution in $\mathcal M_{kE}$ appears, from the viewpoint of observers in $\mathcal M_{xt}$, as a homogeneous, isotropic, and stationary energy density. These are precisely the defining physical properties of a cosmological constant, or dark energy.

It is important to distinguish the present construction from the vacuum energy that appears within the space–time manifold itself.
In standard quantum field theory, vacuum fluctuations of the $(x-t)$ fields arise from local operators in the $(x-t)$ Lagrangian and can therefore be cancelled by renormalization counterterms. By contrast, the stationary ensemble defined on $\mathcal{M}_{kE}$ is a legitimate quantum state of the momentum–energy manifold, specified by its spectral weight $A(k,E)$ and random phases.
This ensemble does not correspond to a local vacuum term of the $(x-t)$ fields.
Its contribution appears in the $(x-t)$ manifold only after the unitary Fourier projection from $\mathcal{M}_{kE}$ to $\mathcal{M}_{xt}$. Because this projected density originates from a nonlocal transformation of degrees of freedom external to the $(x-t)$ Lagrangian, it does not fall within the renormalizable sector of the $(x-t)$ theory and therefore cannot be removed by counterterms.

A potential concern is that $\rho_\Lambda$ might be as large as the quartically divergent vacuum energy of standard quantum field theory~\cite{Adler1995}. In the dual manifold framework this problem does not arise. First, the integrand is the spectral weight $|A|^2$, not the zero-point energy $\frac{1}{2}\hbar\omega_k$; hence the usual $k^3$ ultraviolet divergence is absent. Second, unitarity of the dual-manifold map requires $\int|A|^2<\infty$, which enforces strong suppression of high-$k$ and high-$E$ modes and guarantees that $\rho_\Lambda$ is finite. Third, stationarity on $\mathcal M_{kE}$ implies a finite coherence scale, so the background necessarily exhibits  natural infrared cutoffs $k_0$ and $E_0$ determined by its intrinsic dynamics. What remains undetermined is the overall normalization of $|A(k,E)|^2$, which plays the role of a macroscopic parameter. As in general relativity, the magnitude of the effective cosmological constant is therefore an empirical quantity rather than one derived at this stage of the theory.

\subsection{Black-Hole Horizon and Exponential Boundary Mapping}

Beyond the purely unitary Fourier correspondence, the two manifolds can exhibit dynamical couplings.
Because $\mathcal{M}_{xt}$ and $\mathcal{M}_{kE}$ are Fourier duals, their extreme limits are mapped onto each other: a localized singularity in one manifold corresponds to the asymptotic infinity of the other, and vice versa (Fig. 1). A black hole provides the most direct physical realization of this correspondence. A curvature singularity in $\mathcal{M}_{xt}$ is mapped, under the dual structure, to the asymptotic region of $\mathcal{M}_{kE}$. The event horizon marks the boundary interface where the two manifolds couple: information that becomes inaccessible to observers restricted to $\mathcal{M}_{xt}$ is transferred into modes that propagate toward the asymptotic domain of $\mathcal{M}_{kE}$.

To analyze the effect of this coupling, we examine the behavior of momentum–energy waves as one approaches the asymptotic limit within $\mathcal{M}_{kE}$. Far from the boundary, the coordinates $k$ and $x$ are completely independent: changing $k$ does not affect $x$. An infinitesimal momentum shift $k \to k + \Delta k$ then corresponds to the standard translation–phase relation,
\[
\phi(k-\Delta k) = \int \psi(x)e^{-i\Delta k x} e^{i k x}dx,
\]
so that in $\mathcal{M}_{xt}$ the effect of a momentum translation is merely a linear phase change,
\[
\psi(x) \rightarrow e^{-i \Delta k x}\psi(x).
\]

\begin{figure}[t!]
    \includegraphics[width=8cm]{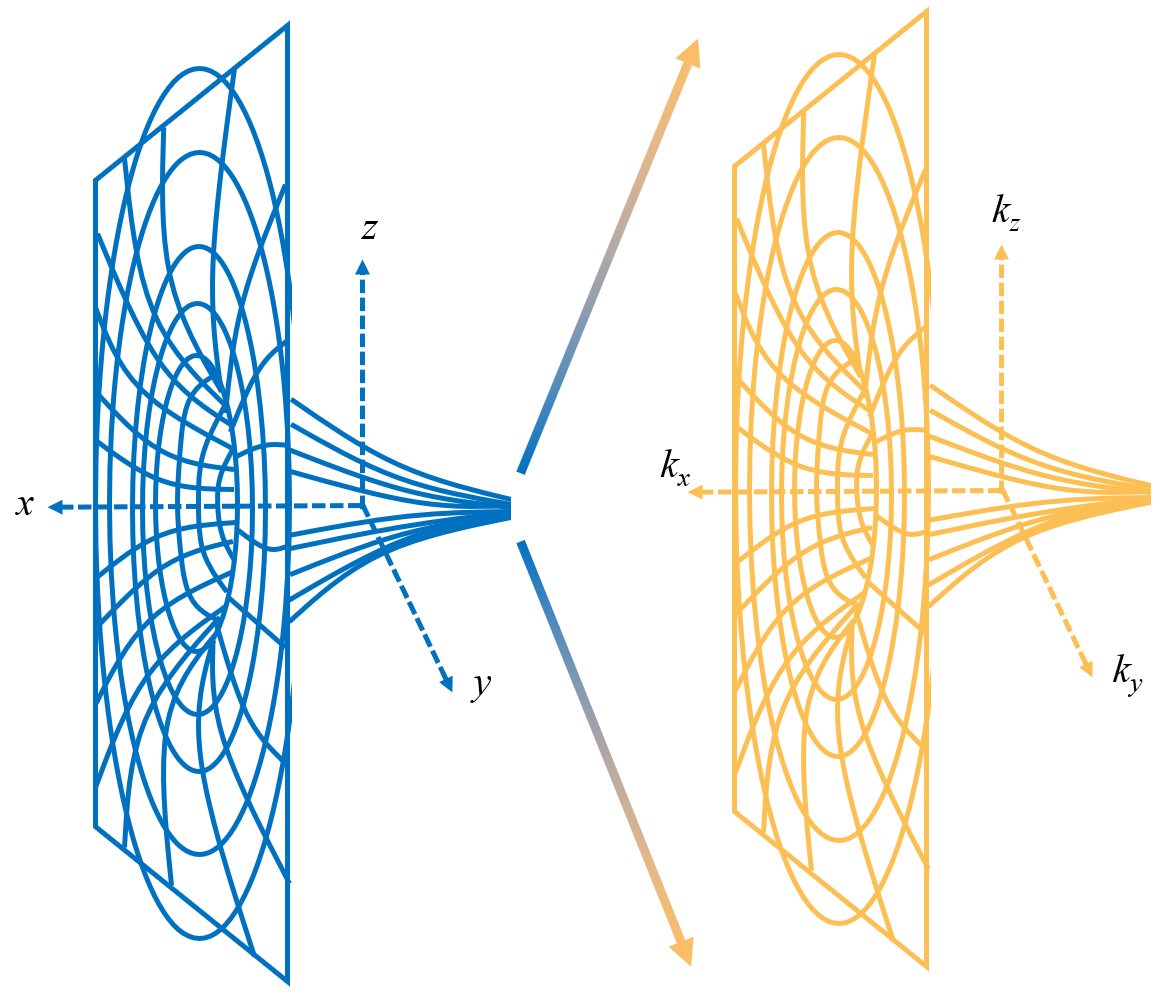}
    \caption{The singularity in the space–time manifold $\mathcal{M}_{xt}$ corresponds to an infinity in the momentum–energy manifold $\mathcal{M}_{kE}$.} 
\end{figure}

As we move toward the boundary $k\to\infty$, however, absolute intervals such as $\Delta k$ lose meaning because any finite additive shift becomes negligible compared to the magnitude of $k$ itself. Only relative changes—ratios between nearby momenta—retain physical significance. Hence, the natural transformation becomes a multiplicative scaling $k \to \lambda k$, where $\lambda > 0$ encodes the local dilation of the momentum coordinate.

Put differently in symmetry language, the transition can be viewed as a change of the fundamental symmetry in the momentum domain—from additive translation 
\[ 
T_{\Delta k}: k \mapsto k + \Delta k, 
\] 
to multiplicative scaling 
\[ 
S_\lambda: k \mapsto \lambda k. 
\]

If we attempt to express this scaling shift in the same additive form, setting $e^{-i \lambda k x} = e^{-i (k - \Delta k)x}$, we find that
\[
\Delta k = k(1 - \lambda),
\]
which makes the momentum shift itself depend on $k$. The standard Fourier kernel $e^{ikx}$ encodes linear translation symmetry, where a shift $k\to k+\Delta k$ acts as a uniform phase factor $e^{-i\Delta k x}$ in the $x$-representation. Once $\Delta k$ depends on $k$, this linear translation symmetry is lost, meaning that the transformation can no longer be realized while holding $x$ fixed. 

To restore consistency with Fourier duality near the boundary, one must therefore allow $x$ to vary as well:
\[
e^{-i\lambda k x} = e^{-i k(x - \Delta x)}.
\]
This indicates that, near the boundary, physical quantities lose their fixed additive scale and become governed by local self-similarity. Each fractional change in momentum produces the same infinitesimal spatial translation $\mathrm dx$. This can be written as
\[
\frac{\mathrm d k}{k} = -\kappa\mathrm d x
\quad\Longleftrightarrow\quad
\frac{\mathrm d\ln k}{\mathrm d x} = -\kappa,
\]
where the constant $\kappa$ denotes the local exponential scaling rate at the boundary. Integrating the differential form gives the exponential map
\[
\ln k = -\kappa x + C
\quad\Rightarrow\quad
k(x) = k_0e^{-\kappa x}.
\]
This exponential correspondence encapsulates the intrinsic coupling of the two manifolds at the boundary, where $k$ and $x$ become coupled variables.

The exponential decay of $k$ with $x$ near the boundary implies that the wave can no longer be expressed in the simple form $Ae^{ikx}$, since the local wavenumber $k$ now varies with $x$. Integrating the phase instead gives
\[
\psi(x) \sim e^{i \int k(x) dx} \sim e^{-iA e^{-\kappa x}},
\]
with a constant $A = k_0/\kappa>0$.

It is important to emphasize that the results obtained here cannot arise within standard quantum mechanics. The key ingredient is the effective symmetry changing from additive translation to multiplicative scaling as one approaches the asymptotic region of $\mathcal M_{kE}$. In ordinary quantum mechanics, the Fourier transform does not require scaling in $k$-space to have any specific counterpart in $\mathcal M_{xt}$.

However, in the dual-manifold framework, $\mathcal M_{kE}$ carries its own autonomous dynamics, making $\mathcal H_{kE}$ a distinct dynamical sector alongside $\mathcal H_{xt}$. The global quantum state possesses simultaneous projections into both sectors, and these projections must remain consistent at their shared boundary. Consequently, the scaling symmetry that naturally emerges in the asymptotic region of $\mathcal M_{kE}$ must be represented coherently in $\mathcal M_{xt}$ sector as well. This boundary-consistency requirement is precisely what forces the exponential relation between $k$ and $x$.

%Although we derived the relation by approaching the asymptotic boundary in $\mathcal{M}_{kE}$, approaching the same interface from $\mathcal{M}_{xt}$ side (toward its singular limit) produces the identical law. Thus the exponential map is not a property of one manifold in isolation; it encodes how the two manifolds project onto each other at the interface.

Strikingly, the exponential boundary relation and the resulting wave structure are mathematically identical to those obtained in the general-relativistic analysis of black-hole horizons. In general relativity, the emergence of such an exponential map is not generic—it relies on the highly constrained geometry near a nonextremal horizon, where the spacetime becomes locally Rindler-like \cite{Rindler1960,Rindler1966}. In that context, an affine coordinate across the horizon (e.g., the Kruskal variable $U$) is exponentially related to the exterior null coordinate $u$,
\[
U=-e^{-\kappa u},
\]
with $\kappa$ denoting the surface gravity. This relation underpins the Hawking effect and other horizon phenomena \cite{Bekenstein1972,HAWKING1974}.

By contrast, in our formulation the same exponential law emerges solely from the dual coupling between $\mathcal M_{xt}$ and $\mathcal M_{kE}$—without any assumption about curvature, metric, or gravitational dynamics. The fact that an identical structure arises purely from restoring the fundamental symmetry between $\mathcal{M}_{xt}$ and $\mathcal{M}_{kE}$ suggests that the exponential map may not be an accidental mathematical coincidence, but a universal feature of boundary geometry itself. It indicates that the exponential relation transcends the specific machinery of general relativity and instead reflects a deeper, symmetry-originated principle.

\section{Conclusion}

To summarize, by endowing the momentum–energy representation with its own autonomous unitary evolution, we restore the intrinsic symmetry between the space–time and momentum–energy manifolds. The momentum–energy domain is not a new addition to quantum mechanics but an inherent part of its Fourier structure; what is new here is the activation of this previously frozen representation into a genuine dynamical sector of the enlarged Hilbert space.
In standard quantum theory, the $(k,E)$ space serves only as a static label set for spectral decomposition, while all evolution takes place in $(x,t)$.
Here, the momentum–energy representation acquires its own generator of evolution, transforming a passive representation into an active physical manifold capable of dynamically coupling back to $(x,t)$.
This restored symmetry provides a more complete and balanced description of quantum waves and allows many phenomena—previously described through independent assumptions—to emerge from a single coherent framework. As discussed above, for instance, the dark energy and the exponential relation underlying Hawking radiation follows directly from the projection between the two manifolds, without invoking general relativity.

Other long-standing puzzles may be reinterpreted in the same light.
For example, the black-hole information paradox finds a natural resolution once both dynamical sectors of the global Hilbert space are considered: information is not destroyed inside the black hole but transitions into its conjugate sector, where propagation within $\mathcal{M}_{kE}$ becomes inaccessible to observers confined to $\mathcal{M}_{xt}$.

%In this work we focused on boundary couplings—interactions mediated across the shared projection interface between the two manifolds. More generally, however, the coupling need not be restricted to this boundary. Non-boundary interactions, in which excitations in one manifold project indirectly into the other without crossing their common interface, could give rise to hidden dynamical sectors. Such phenomena may manifest observationally as dark matter: degrees of freedom residing primarily in the conjugate manifold but weakly entangled with our own.

Taken together, the dual-manifold framework suggests a new, self-consistent picture of physical reality: a universe described by a single quantum state with two complementary dynamical manifestations, where neither manifold is complete on its own, but their interplay defines the full structure of existence. We hope that further investigations will expand upon this foundation and open a new chapter in the pursuit of fundamental physics.

\bibliographystyle{unsrt}
\bibliography{kspace}

\clearpage

\end{document}